\documentclass{PoS}

\usepackage{amsfonts,amssymb,amsmath,bm}
\usepackage{graphicx}

\title{The effective gluon mass \\ and its dynamical equation.}

\ShortTitle{The effective gluon mass and its dynamical equation.}

\author{\speaker{Joannis Papavassiliou}\\     
        Department of Theoretical Physics and IFIC,\\ 
        University of Valencia-CSIC,\\
        E-46100, Valencia, Spain.\\
        E-mail: \email{Joannis.Papavassiliou@uv.es}}

\author{David Ibanez\\
        Department of Theoretical Physics and IFIC,\\
        University of Valencia-CSIC,\\
        E-46100, Valencia, Spain.\\
        E-mail: \email{daigilde@alumni.uv.es}}

\abstract{We present the general derivation of the full nonperturbative equation  
that governs the momentum evolution of the dynamically generated gluon  
mass, in the Landau gauge. The gluon mass originates from the inclusion of  
longitudinally coupled vertices containing massless poles of  
non-perturbative origin, which preserve the form of the fundamental  
Slavnov-Taylor identities of the theory. The equation is obtained within  
the PT-BFM formalism, where the corresponding Schwinger-Dyson equation  
involves a reduced number of fully dressed diagrams. The resulting  
homogeneous integral equation is solved numerically for the entire range  
of physical momenta, yielding positive-definite and monotonically  
decreasing gluon masses, in agreement with a variety of less formal  
considerations.}

\FullConference{$X_{th}$ Quark Confinement and the Hadron Spectrum\\
		 7-12 October 2012\\
		 M\"unchen, Germany}

\begin{document}

\section{Introduction}

It is well-established by now that the dynamical generation of an effective 
gluon mass~\cite{Cornwall:1981zr} explains in a natural and self-consistent 
way the infrared finiteness of the (Landau gauge) gluon propagator and ghost 
dressing function, observed in large-volume lattice simulations for both 
$SU(2)$~\cite{Cucchieri:2007md} and $SU(3)$ gauge groups~\cite{Bogolubsky:2007ud}. 
Given the nonperturbative nature of the mass generation mechanism, the 
Schwinger-Dyson equations (SDEs) constitute the most natural framework for 
studying such phenomenon in the continuum~\cite{Binosi:2007pi,Alkofer:2000wg,Fischer:2006ub,Aguilar:2004sw}. 
Specifically, we will work in the framework provided by the synthesis of the 
pinch technique (PT)~\cite{Cornwall:1981zr,Cornwall:1989gv,Binosi:2002ft}
 with the background field method (BFM)~\cite{Abbott:1980hw}, known in the 
 literature as the PT-BFM scheme~\cite{Binosi:2002ft,Aguilar:2006gr}.

Probably the most crucial theoretical ingredient for obtaining out of the SDEs 
an infrared-finite gluon propagator, \textit{without} interfering with the gauge invariance 
of the theory, encoded in the BRST symmetry, is the existence of a set of special 
vertices, to be generically denoted by $V$ and called \textit{pole vertices}. These vertices 
contain massless, longitudinally coupled poles, and must be added to the usual (fully dressed) 
vertices of the theory. They capture the underlying mass generation mechanism, which is 
none other than a non-Abelian realization of the Schwinger mechanism. In addition 
to triggering the Schwinger mechanism, the massless poles contained in the pole vertices act 
as composite, longitudinally coupled Nambu-Goldstone bosons, maintaining gauge invariance and 
preserving the form of the Ward identities (WIs) and the Slavnov-Taylor identities 
(STIs) of the theory in the presence of a dynamically generated gluon mass. 
In fact, recent studies indicate that the QCD dynamics can indeed generate longitudinally 
coupled composite (bound-state) massless poles, which subsequently give raise to the required 
vertices $V$~\cite{Aguilar:2011xe,Ibanez:2012zk}.

At the level of the SDEs, the analysis finally boils down to the derivation of an 
integral equation, to be referred as the \textit{mass equation}, that governs the evolution 
of the dynamical gluon mass, $m^2(q^2)$, as a function of the momentum $q^2$. The 
main purpose of this presentation is to report on recent work~\cite{Binosi:2012sj}, 
where the complete mass equation has been obtained in the Landau gauge employing the 
\textit{full} SDE of the gluon propagator and using as a guiding principle the special properties 
of the aforementioned vertices $V$ (for related studies in the Coulomb gauge see, 
\textit{e.g.},~\cite{Szczepaniak:2001rg,Epple:2007ut}). 
In this context, the detailed numerical solution 
of the full mass equation (for arbitrary values of the physical momentum), reveals 
the existence of \textit{positive-definite} and \textit{monotonically decreasing} 
solutions.

\section{The SDE of the gluon propagator}
\begin{figure}[t!]
\center{\includegraphics[scale=0.43]{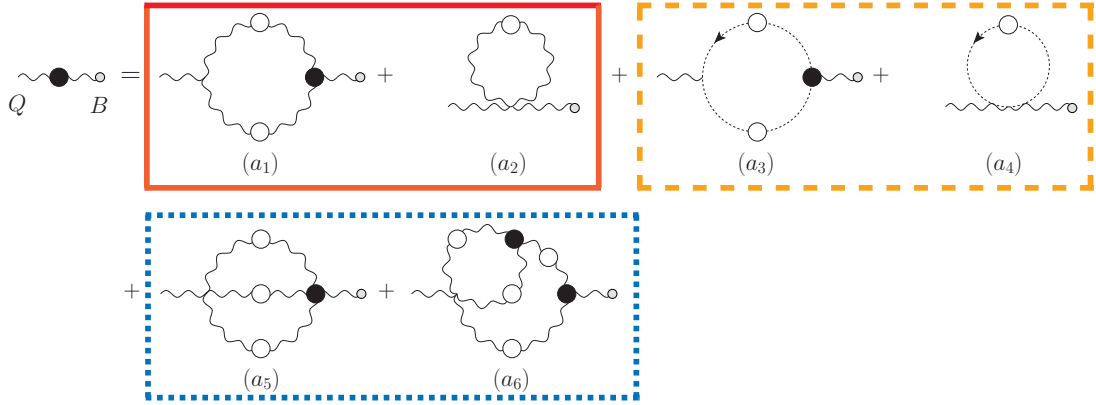}}
\caption{The SDE obeyed by the $QB$ gluon propagator. Black blobs represents fully 
dressed 1-PI vertices; the small gray circles appearing on the external legs (entering 
from the right, only!) are used to indicate background gluons.}
\label{QB-SDE}
\end{figure}
The full gluon propagator $\Delta^{ab}_{\mu\nu}(q)=\delta^{ab}\Delta_{\mu\nu}(q)$ 
in the Landau gauge is given by the expression
\begin{equation}\label{prop}
\Delta_{\mu\nu}(q)=-iP_{\mu\nu}(q)\Delta(q^2); \quad P_{\mu\nu}(q)=g_{\mu\nu}-\frac{q_\mu q_\nu}{q^2},
\end{equation}
and its \textit{inverse} gluon dressing function, $J(q^2)$, is defined as
\begin{equation}\label{gldressing}
\Delta^{-1}(q^2)=q^2 J(q^2).
\end{equation}
The usual starting point of our dynamical analysis is the SDE governing the gluon propagator. 
Specifically, within the PT-BFM formalism, one can consider the propagator connecting a quantum 
($Q$) with a background ($B$) gluon, to be referred as the $QB$ propagator and denoted by 
$\widetilde{\Delta}(q^2)$. The SDE of the above propagator is shown in Fig.~\ref{QB-SDE}, and 
it may be related to the conventional $QQ$ propagator, $\Delta(q^2)$, connecting two quantum gluons, 
through the powerful background-quantum identity~\cite{Binosi:2002ft,Grassi:1999tp}
\begin{equation}\label{propBQI}
\Delta(q^2) = [1+G(q^2)]\widetilde{\Delta}(q^2),
\end{equation}
In this identity, the function $G(q^2)$ corresponds to the $g_{\mu\nu}$ form factor of a well known 
two-point function~\cite{Binosi:2007pi,Binosi:2002ft,Grassi:1999tp}. Then, the corresponding version 
of the SDE for the conventional gluon propagator (in the Landau gauge) 
reads~\cite{Binosi:2007pi,Aguilar:2006gr}
\begin{equation}\label{glSDE}
\Delta^{-1}(q^2)P_{\mu\nu}(q) = \frac{q^2P_{\mu\nu}(q) + i\sum_{i=1}^6(a_i)_{\mu\nu}}{1 + G(q^2)},
\end{equation}
where the diagrams $(a_i)$ are shown in Fig.~\ref{QB-SDE}. The relevant point to recognize here is 
that the transversality of the gluon self-energy is realized according to the pattern highlighted by 
the boxes of Fig.~\ref{QB-SDE}, namely,
\begin{equation}\label{pattern}
q^\mu[(a_1)+(a_2)]_{\mu\nu}=0; \quad q^\mu[(a_3)+(a_4)]_{\mu\nu}=0; \quad q^\mu[(a_5)+(a_6)]_{\mu\nu}=0.
\end{equation}

\section{Derivation of the gluon mass equation}
As has been explained in detail in the recent literature~\cite{Aguilar:2011xe,Ibanez:2012zk}, 
the Schwinger mechanism allows for the emergence of massive solutions out of the SDE, preserving,  
at the same time, the gauge invariance intact. At this level, the triggering of this mechanism 
proceeds through the inclusion of the pole vertices $V$ in the SDE Eq.~(\ref{glSDE}). From the 
kinematic point of view, we will describe the transition from a massless to a massive gluon 
propagator by carrying out the replacement (Minkowski space)
\begin{equation}\label{massiveprop}
\Delta^{-1}(q^2)=q^2 J(q^2) \longrightarrow \Delta_m^{-1}(q^2) = q^2 J_m(q^2) - m^2(q^2).
\end{equation}
Notice that the subscript ``m'' indicates that effectively one has now a mass inside the corresponding 
expressions: for example, whereas perturbatively $J(q^2)\sim \ln q^2$, after dynamical gluon mass 
generation has taken place, one has $J_m(q^2)\sim \ln(q^2 + m^2)$. Then, gauge invariance requires 
that the replacement given in Eq.~(\ref{massiveprop}) be accompanied by the following simultaneous 
replacement of all relevant vertices
\begin{equation}\label{replacever}
\Gamma \longrightarrow \Gamma' = \Gamma_m + V,
\end{equation}
where $V$ must be such that the new vertex $\Gamma'$ satisfies the same formal WIs (or STIs) as 
$\Gamma$ before. The most familiar case is that of the $BQ^2$ vertex 
$\widetilde{\Gamma}'_{\alpha\mu\nu}$, whose pole part must satisfy the WI~\cite{Aguilar:2011ux} 
\begin{equation}\label{WIthreepole}
q^\alpha \widetilde{V}_{\alpha\mu\nu}(q,r,p) = m^2(r^2) P_{\mu\nu}(r) - m^2(p^2) P_{\mu\nu}(p),
\end{equation}
when contracted with respect to the momentum of the background gluon. In complete analogy with the 
above case, one may use the WI satisfied by the conventional $BQ^3$ vertex, namely,
\begin{eqnarray}\label{WIfour}
q^\alpha \widetilde{\Gamma}_{\alpha\mu\nu\rho}^{abcd}(q,r,p,t) &=& ig^2[f^{abx}f^{xcd}\Gamma_{\nu\rho\mu}(p,t,q+r) + f^{acx}f^{xdb}\Gamma_{\rho\mu\nu}(t,r,q+p) \nonumber \\
&+& f^{adx}f^{xbc}\Gamma_{\mu\nu\rho}(r,p,q+t)],
\end{eqnarray}
in order to deduce that, after the replacement Eq.~(\ref{replacever}), its $\widetilde{V}$ part 
satisfies~\cite{Binosi:2012sj}
\begin{eqnarray}\label{WIfourpole}
q^\alpha \widetilde{V}_{\alpha\mu\nu\rho}^{abcd}(q,r,p,t) &=& ig^2[f^{abx}f^{xcd}V_{\nu\rho\mu}(p,t,q+r) + f^{acx}f^{xdb}V_{\rho\mu\nu}(t,r,q+p) \nonumber \\
&+& f^{adx}f^{xbc}V_{\mu\nu\rho}(r,p,q+t)].
\end{eqnarray}
Finally, as a large variety of lattice simulations and analytic studies suggest, 
we will take for granted that the ghost propagator $D(q^2)$ remains massless in the Landau gauge. 
The main implication of this property for the case at hand is that the (fully-dressed) BFM gluon-ghost vertex, 
appearing in graph $(a_3)$, does \textit{not} need to be modified by the presence of 
$\widetilde{V}$-type vertices.

Quite remarkably, the above WIs, supplemented 
by the totally longitudinal nature of the pole vertices, are the only properties that one needs for deriving the mass 
equation; in particular, the closed form of the pole vertices is not needed.

According to the previous discussion, after the inclusion of the pole vertices, 
the gluon SDE Eq.~(\ref{glSDE}) becomes in the Landau gauge
\begin{equation}\label{glSDEprime}
[q^2J_m(q^2) - m^2(q^2)]P_{\mu\nu}(q) = \frac{q^2P_{\mu\nu}(q) + i\sum_{i=1}^6(a'_i)_{\mu\nu}}{1 + G(q^2)},
\end{equation}
where the \textit{prime} indicates that (in general) one must perform the simultaneous 
replacements Eq.~(\ref{massiveprop}) and Eq.~(\ref{replacever}) inside the corresponding 
diagrams. Evidently, the lhs of Eq.~(\ref{glSDEprime}) involves two unknown quantities, 
$J_m(q^2)$ and $m^2(q^2)$, which will eventually satisfy two separate, but \textit{coupled}, 
integral equations of the generic type
\begin{eqnarray}
&& J_m(q^2) = 1 + \int_k {\cal K}_1(q^2,m^2,\Delta_m), 
\nonumber \\
&& m^2(q^2) = \int_k {\cal K}_2(q^2,m^2,\Delta_m), 
\label{meq}
\end{eqnarray}
such that ${\cal K}_1, {\cal K}_2\neq 0$, as $q^2\rightarrow 0$. In order to derive the closed form 
of the mass equation Eq.~(\ref{meq}), one must identify all mass-related contributions coming from 
the $\widetilde{V}$ vertices that are contained in the Feynman graphs comprising the rhs of 
Eq.~(\ref{glSDEprime}). With the transversality of both sides of Eq.~(\ref{glSDEprime}) guaranteed 
by the presence of the pole vertices, it is far more economical to derive the mass equation by isolating 
the appropriate cofactors of $q_\mu q_\nu/q^2$, to be denoted by $a_i^{\widetilde{V}}(q^2)$, on both sides. 
Notice that selecting the $g_{\mu\nu}$, or taking the trace in Eq.~(\ref{glSDEprime}), would entail the 
use of the special \textit{seagull identity}~\cite{Aguilar:2009ke,Aguilar:2011ux}. 

The most important steps of this construction may   
be summarized as follows: $({\bf i})$ From the previous comments about the BFM gluon-ghost vertex, graph 
$(a'_3)$ have not $\widetilde{V}$-component. $({\bf ii})$ The WI Eq.~(\ref{WIfourpole}) and the longitudinality 
condition for the $BQ^3$ pole vertex may be used to demonstrate that the $\widetilde{V}$-component of graph 
$(a'_5)$ vanishes in the Landau gauge. $({\bf iii}) $ The contribution $a_6^{\widetilde{V}}(q^2)$ stems 
solely from the combination $\Gamma_m\widetilde{V}$ of the product $\Gamma'\widetilde{\Gamma}'$ appearing 
in graph $(a'_6)$. 

\begin{figure}[t!]
\center{\includegraphics[scale=0.55]{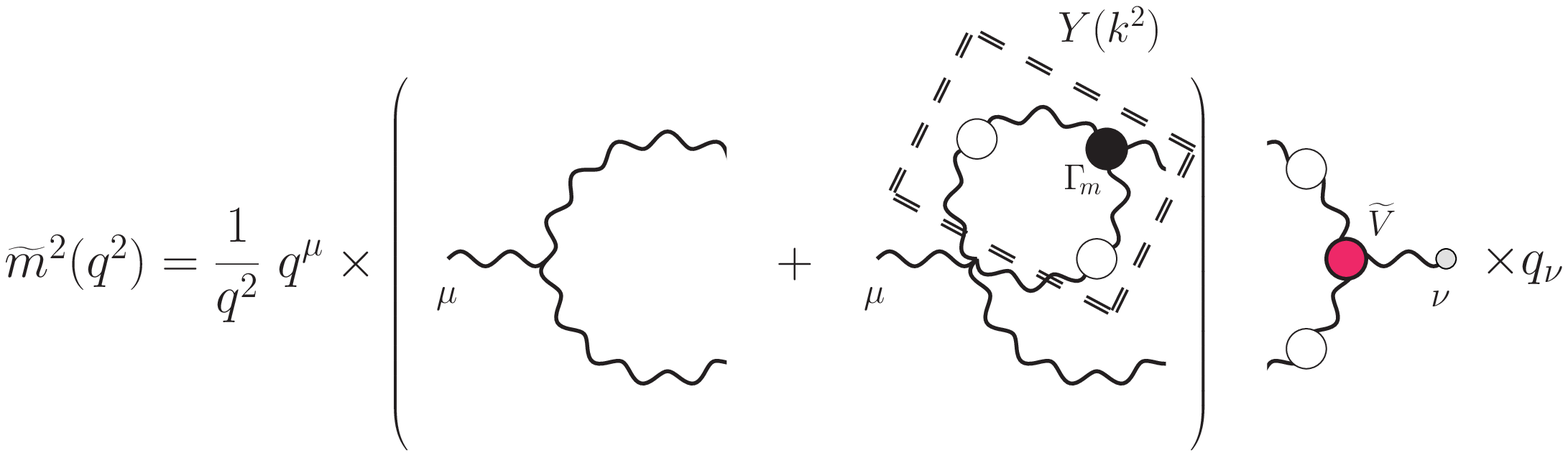}}
\caption{Diagrammatic representation of the condensed operations leading to 
the all-order gluon mass equation, where we have introduced the shorthand 
notation $\widetilde{m}^2(q^2)=m^2(q^2)[1+G(q^2)]$. All internal propagators
 are in the Landau gauge.}
\label{diagrammaticmass}
\end{figure}

Thus, one concludes that the \textit{complete} mass equation can be written as 
\begin{equation}\label{masscompact}
m^2(q^2) = \frac{i[a_1^{\widetilde{V}}(q^2) + a_6^{\widetilde{V}}(q^2)]}{1 + G(q^2)}.
\end{equation}
Interestingly enough, the entire procedure may be pictorially summarized, in a rather concise way, as shown in Fig.~\ref{diagrammaticmass}.

\section{Complete mass equation and numerical results}

The final equation obtained from Eq.~(\ref{masscompact}) reads (Euclidean space)
\begin{eqnarray}\label{masscomplete}
m^2(q^2) &=& -\frac{g^2C_A}{1+G(q^2)}\frac{1}{q^2}\int_k m^2(k^2)[(k+q)^2-k^2]
\Delta^{\alpha\rho}(k)\Delta_{\alpha\rho}(k+q)\bigg\lbrace 1 - C\,[Y(k+q)+Y(k)]\bigg\rbrace \nonumber \\
&+& \frac{g^2C_A}{1+G(q^2)}\frac{1}{q^2}(q^2g_{\delta\gamma}-2q_\delta q_\gamma)\int_k m^2(k^2)\, C\, [Y(k+q)-Y(k)]
\Delta_\epsilon^\delta(k)\Delta^{\gamma\epsilon}(k+q),
\end{eqnarray}
with  $C=3\pi C_A\alpha_s$, and 
\begin{equation}\label{Yintegral}
Y(k^2)=\frac{1}{3}\frac{k_\alpha}{k^2}\int_l \Delta^{\alpha\rho}(l)\Delta^{\beta\sigma}(l+k)\Gamma_{\sigma\rho\beta},
\end{equation}
corresponding to the subdiagram on the upper left corner of $(a_6)$ [see also Fig.~\ref{diagrammaticmass}].

Even though Eq.~(\ref{masscomplete})
forms part of a system of coupled equations, see Eq.~(\ref{meq}), in what follows we will study it in isolation, given that the 
corresponding equation for $J_m (q^2)$ is unknown. To that end, we will treat the gluon propagators appearing 
in the mass equation as external quantities, using lattice results for their form~\cite{Bogolubsky:2007ud}.   

In addition, the rhs of Eq.~(\ref{Yintegral}) 
depends on the full three-gluon vertex $\Gamma_{\sigma\rho\beta}$, whose exact form is not known.
We will therefore approximate $Y(k^2)$ by its one-loop expression, obtained by substituting tree-level values for  
all quantities appearing in the integral; a lengthy but straightforward 
calculation yields [Euclidean space, momentum subtraction (MOM) scheme]
\begin{equation}\label{Yrenor}
Y_R(k^2)=-\frac{1}{(4\pi)^2}\frac{5}{4}\log\frac{k^2}{{\mu}^2}\,.
\end{equation}

After these considerations, and using spherical coordinates $x=q^2$ and 
$y=k^2$, let us study the deep infrared limit $x\rightarrow 0$ of Eq.~(\ref{masscomplete}), 
given by
\begin{eqnarray}\label{deepIR}
&& m^2(0) = -\frac{3\alpha_S C_A}{8\pi [1+G(0)]}\int_0^\infty dy  m^2(y){\cal K}_2(y); \nonumber \\
&& {\cal K}_2(y) = \lbrace[1-2CY(y)]Z^2(y)\rbrace', \quad Z(y)=y\Delta(y).
\end{eqnarray}

Even though the value of $C$ is fixed (see above), in what follows we will treat it 
as a free parameter, in order to study 
what happens to the gluon mass equation when one varies independently $\alpha_S$ and $C$. 
The reason for doing this is that, whereas Eq.~(\ref{Yrenor}) furnishes a concrete form for the two-loop \textit{dressed} correction, 
by no means does it exhaust it; thus, by varying $C$, one basically tries to mimic further correction that 
may be added to the \textit{skeleton} provided by the $Y_R(k^2)$ of Eq.~(\ref{Yrenor}) (for a fixed value of $\alpha_S$); 
indeed rescaling C is equivalent to rescaling $Y_R(k^2)$.
Of course, the real value of $C$ will emerge as a special case of this general two-parameter study. 
Finally, it is convenient to define the ``reduced'' $C_r=C/3\pi C_A$, and drop the suffix ``r''.
\begin{figure}[t!]
\center{\includegraphics[scale=0.8]{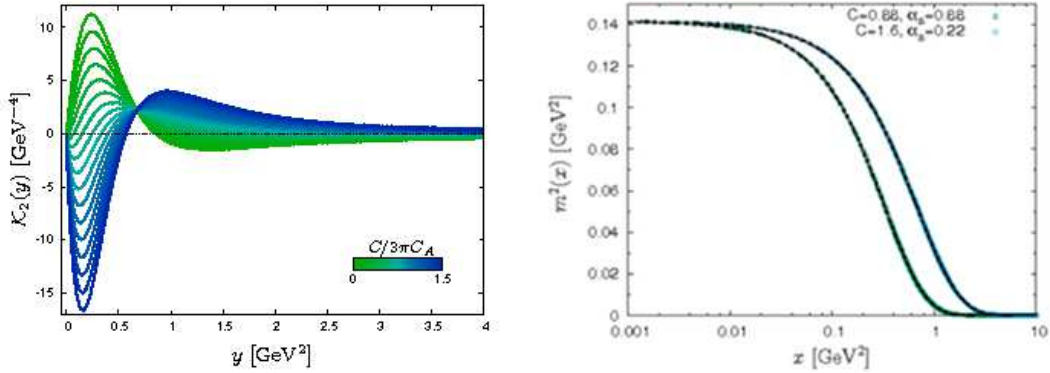}}
\caption{{\it Left panel}: Modification in the shape of the two-loop dressed kernel ${\cal K}_2(y)$ when  
varying $C$. As $C$ increases, the kernel effectively reverses its sign, showing a deep 
negative well in the low momenta region. {\it Right panel}: Typical monotonically decreasing solution 
of Eq. (5.1). The case shown has been obtained for the special values $C=0.88$ and $1.85$, corresponding 
value of the coupling $\alpha_S\approx 0.88$ and $0.22$, respectively. The solutions have been normalized 
so that at the origin they match the corresponding (Landau gauge) lattice value $\Delta^{-1}(0)\approx 0.141$ 
GeV$^2$, namely $m(0)=375$ MeV.}
\label{kernel_mass}
\end{figure}

Let us first set $C=0$, thus turning off the two-loop dressed contributions. Then, integrating (\ref{deepIR}) by parts, one obtains 
\begin{equation}
m^2(0) = \frac{3\alpha_S C_A}{8\pi [1+G(0)]}\int_0^\infty dy  [m^2(y)]' Z^2(y)\,.
\end{equation}
Given that $1+G(0)>0$, 
it is clear that a monotonically decreasing gluon mass, namely \mbox{$[m^2(y)]' <0$}, expected on physical grounds, 
would give rise to a negative value for $m^2(0)$, which is physically wrong.
Thus, the only way to reconcile a positive-definite and monotonically decreasing gluon mass is to 
obtain an effective reversal of sign from the two-loop dressed contributions; as we will see, this is indeed what happens. 

To study this crucial point in detail, 
let us consider how the shape of the kernel ${\cal K}_2$ changes as $C$ is varied. In 
Fig.~\ref{kernel_mass}, one observes that, as $C$ increases, ${\cal K}_2$ displays a less pronounced 
positive (respectively negative) peak in the small (respectively large) momenta region. Next, for 
$C\gtrsim 0.37$, a small negative region starts to appear in the infrared, which rapidly becomes a deep 
negative well for $y\lesssim 0.6$, with ${\cal K}_2$ becoming positive for higher momenta. Therefore we 
observe that the addition of the two-loop dressed contributions counteracts the effect of the overall 
minus sign of Eq.~(\ref{deepIR}), by effectively achieving a sign reversal of the kernel. Indeed, one 
concludes that there exists a critical value $\overline{C}\approx 0.56$ such that, if $C>\overline{C}$, 
Eq.~(\ref{deepIR}) will display at least one physical monotonically decreasing solution for a suitable 
value of the strong coupling $\alpha_S$.

Finally, to see if the picture sketched above is confirmed when $x\neq 0$, one can study numerically the 
solutions of Eq.~(\ref{masscomplete}) following the algorithm described in~\cite{Binosi:2012sj}. In this 
case the absence of solutions persists until the critical value $\overline{C}$ is reached, after which 
one finds exactly one monotonically decreasing solution. Specifically, in Fig.~\ref{kernel_mass} we plot 
the solutions for the most representative $C$ values. The value $C=\alpha_S\approx 0.88$ corresponds to 
the case in which $Y$ is kept at its lowest order perturbative value, whereas $C=1.85$ 
corresponds to the standard MOM value $\alpha_S=0.22$ at $\mu=4.3$ GeV~\cite{Boucaud:2005rm}.
As can be readily appreciated, the masses 
obtained display the basic qualitative features expected on general field-theoretic considerations and 
employed in numerous phenomenological studies; in particular, they are monotonically decreasing functions 
of the momentum and vanish rather rapidly in the ultraviolet.

\section{Conclusions}
In this presentation we have reported recent progress~\cite{Binosi:2012sj} on the study 
of the nonpertubative equation that governs the momentum evolution of the dynamically 
generated gluon mass. By appealing to the existence of the special nonperturbative vertices 
$V$ associated with the Schwinger mechanism, we have outlined the methodology that allows 
for a systematic and expeditious identification of the parts of the SDE that contributes 
to the mass equation. The numerical analysis of the resulting mass equation reveals that 
the inclusion of two-loop dressed contributions has a profound impact on the nature of the 
mass equation, already at the qualitative level. Indeed, they are crucial in order to obtain 
physically meaningful solutions out of the mass equation, \textit{i.e.}, positive-definite 
and monotonically decreasing solutions for the effective gluon mass. 

In the future, given the importance of the term $Y(k^2)$ for this entire construction, it would 
be particularly important to determine its structure beyond the perturbative one-loop approximation used. 
Nevertheless, even with the approximate version of $Y(k^2)$, the full mass equation provides a 
natural starting point for calculating reliably the effect that the inclusion of light quark 
flavors might have on the form of the gluon propagator, and complement recent studies based on 
the SDEs~\cite{Aguilar:2012rz} as well as lattice simulations~\cite{Ayala:2012pb}.

\acknowledgments

This research is supported by the European FEDER and Spanish MICINN under grant FPA2008-02878.

\end{document}